\documentstyle[sprocl]{article}

\def\ii{{\rm i}}
\def\de{{\rm\,d}}
\def\e{{\rm e}}

\def\<{\langle} %bra
\def\>{\rangle} %ket
\def\tr{\mbox{Tr}\,}

\def\eq{\begin{equation}}
\def\feq{\end{equation}}
\def\be{\begin{equation}}
\def\ee{\end{equation}}
\def\ar{\begin{eqnarray}}
\def\far{\end{eqnarray}}
\def\bea{\begin{eqnarray}}
\def\eea{\end{eqnarray}}
\def\^{\hat}

\newcommand{\bm}[1]{\mbox{\boldmath $#1$}}
\newcommand{\g}{\gamma}
\newcommand{\sig}{\sigma}
\newcommand{\eps}{\epsilon}
\newcommand{\nn}{\nonumber}

\newcommand{\Pslash}{\kern 0.2 em P\kern -0.56em \raisebox{0.3ex}{/}}
\newcommand{\Sslash}{\kern 0.2 em S\kern -0.56em \raisebox{0.3ex}{/}}
\newcommand{\pslash}{\kern 0.2 em p\kern -0.4em /}
\newcommand{\kslash}{\kern 0.2 em k\kern -0.45em /}
\newcommand{\qslash}{\kern 0.2 em q\kern -0.4em /}
\newcommand{\nslash}{\kern 0.2 em n\kern -0.5em /}

\def\bold#1{\setbox0=\hbox{$#1$}%
     \kern-.02em\copy0\kern-\wd0
     \kern.04em\copy0\kern-\wd0
     \kern-.02em\raise.0433em\box0 }

\begin{document}

\title{SEMI-INCLUSIVE VECTOR MESON PRODUCTION IN DIS}

\author{\underline{A. Bacchetta}, P.J. Mulders}

\address{Department of Theoretical Physics, Faculty of Science, 
Free University \\
De Boelelaan 1081, NL-1081 HV Amsterdam, the Netherlands}

%%%%%%%%%%%%%%%%%%%%%%%%%%%%%%%%%%%%%%%%%%%%%%%%%%%%%%%%%%%%%%%%%%%%%%%

\maketitle\abstracts{We analyze one-particle inclusive DIS in the case when
a spin-1 hadron (such as a vector meson) is observed in the final state. We
 consider only leading order contributions in $1/Q$, but we include
transverse momentum of partons. Several new
fragmentation functions appear in cross sections. One of them can be measured
in connection with the transverse-spin distribution function $h_1$.}

%%%%%%%%%%%%%%%%%%%%%%%%%%%%%%%%%%%%%%%%%%%%%%%%%%%%%%%%%%%%%%%%%%%%%%%

The spin density matrix of a spin-1 particle \cite{bls} can be decomposed on a
Cartesian basis using the spin vector $S^i$ and a symmetric traceless 
rank-two spin tensor $T^{ij}$ :
\eq
{\bm \rho}=\frac{1}{3}\left({\bf 1} + \frac{3}{2} S^i {\bm\Sigma}^i 
		+ 3\, T^{ij} {\bm\Sigma}^{ij}\right),
\label{e:density}
\feq
where ${\bm\Sigma}$'s form a suitable basis of $3 \times 3$ matrices.
We parametrize the spin vector and tensor
in the rest-frame of the hadron in the following way:
\begin{equation}
{\bm S}=\left(S_{T}^x, S_{T}^y, S_{L}\right),  \qquad
{\bm T}=\frac{1}{2}\left(\begin{array}{ccc}
	-\frac{2}{3}{S_{LL}}+{S_{TT}^{xx}} 
		& {S_{TT}^{xy}}	& {S_{LT}^{x}} \\
	{S_{TT}^{xy}}& 	-\frac{2}{3}{S_{LL}}-{S_{TT}^{xx}}
					& {S_{LT}^{y}}  \\
	 {S_{LT}^{x}}	& {S_{LT}^{y}}	& \frac{4}{3}{S_{LL}} \\
	\end{array}\right).
\label{e:tensor}
\end{equation}

The spin tensor of an outgoing vector meson can be extracted from the angular 
distribution of the decay products, although it is not possible to extract any
information on the spin
vector in an analogous manner. For instance, in the case of a
$\rho$-meson decaying into $\pi^+ \pi^-$, the decay distribution depends only
on the spin tensor components, being \cite{sch}
\begin{eqnarray}
W( \theta, \varphi)
	&=& \frac{3}{8\pi}\left\{\frac{2}{3}
	-\frac{2}{3}\,S_{LL}(\cos^2{\theta}+\cos{2\theta})
	-\sin{2\theta}\; (S_{LT}^{x}\,\cos{\varphi} 
		+S_{LT}^{y} \, \sin{\varphi})\right.  \nn \\ 
&&	\left.{\phantom{\frac{2}{3}}}
	\mbox{}-\sin^2{\theta}\;(S_{TT}^{xx}\, \cos{2\varphi}
	+S_{TT}^{xy}\, \sin{2\varphi})\right\},
\end{eqnarray}
where $\varphi$ is the azimutal angle between the meson production
plane and the pions decay plane, and $\theta$ is the polar angle of one
of the two pions in the $\rho$ rest-frame.

To describe  semi-inclusive DIS we need to introduce two soft correlation 
functions, describing the quark distribution in the spin-1/2 target and 
the hadronization of a quark into the final state spin-1 hadron.  
In leading order in $1/Q$ (also referred to as ``leading twist'' or
``twist-2'') we are  concerned only with the 
quark-quark correlation functions $\Phi$ and $\Delta$. 
The correlation function $\Phi$ has been already
widely studied in the literature (see e.g. \cite{multa}).
The function $\Delta$ is defined as (using
Dirac indices $\alpha$ and $\beta$)
\ar
\Delta_{\alpha\beta}(k,P_h,S_h,T_h)&=&\int
        \frac{\de^{4}\xi}{(2\pi)^{4}}\e^{+\ii k\xi}
        \<0|\psi_{\alpha}(\xi)|P_h,T_h\>\<P_h,T_h|
             \bar{\psi}_{\beta}(0)|0\>.    
\label{e:corr}
\far
Here, $k$ is 
the momentum of the quark decaying into an outgoing hadron after being struck
by a virtual photon. The vector $P_h$
is the momentum of the outgoing hadron, $S_h$ is its spin vector 
and $T_h$ is its spin tensor. 
In the hadronic tensor we need the integrated correlation function
\ar
\Delta(z ,{\bm k}_{T})& =& 
		\frac{1}{4z} \left. \int \de k^+ \;
                    \Delta(k,P_h,S_h,T_h) \right|_{k^-=\frac{P_h^-}{z}\,;\,
						k_T=\frac{P_{h\perp}}{z} }\, , 
\label{e:frag}
\far
which can be decomposed using 18 different fragmentation functions \cite{io}.
 If we do
not observe the perpendicular component of the momentum of the outgoing hadron,
we need to perform a further integration on ${\bm k}_{T}$ and to deal with the
 function
\ar
\Delta(z)& =& \frac{z}{4} \left. \int \de^2 {\bm k}_T \, \de k^+ \;
                    \Delta (k,P_h,S_h,T_h)\right|_{k^-=\frac{P_h^-}{z}}. 
\label{e:frag2}
\far 
The parametrization of the correlation function after integration upon
${\bm k}_T$, as defined in Eq.~(\ref{e:frag2}), is
\begin{eqnarray} 
\Delta (z)& =& \frac{1}{4}\left\{
		   D_1(z)\nslash_-
		+  D_{1LL}(z)\,S_{h LL} \,\nslash_- 
		+  G_{1}(z)\,S_{h L}\,\g_5 \nslash_- \right. \nn \\
& &\left.\mbox{}+  H_{1}(z)\,
		\ii \sig_{\mu \nu} \g_5 n_-^{\mu} S_{h T}^{\nu}
		+  H_{1LT}(z)\,
    \ii \sig_{\mu \nu} \g_5 n_-^{\mu}\eps_T^{\nu \rho}S_{h LT\,\rho}
		\right\},  
\end{eqnarray}
where the fragmentation function $H_{1LT}$ is chiral odd and time-reversal
odd. This function has already been discussed by Ji \cite{ji} where it was
named $\^h_{\bar{1}}$. Since $\rho$ decay into two pions does not allow to
measure the spin vector, 
it is not possible to observe the fragmentation functions
$G_1$ and $H_1$.

The cross-section of semi-inclusive deep-inelastic scattering is
\begin{equation}
\frac{\de \sig (l+H \rightarrow l'+h+X)}
	{\de x \de z \de y}=
	\frac{\pi \alpha^2}{Q^4}\;\frac{y}{2 z}\; L_{\mu \nu}\; 2M W^{\mu \nu},
\label{e:cross}
\end{equation}
where $x$, $z$ and $y$ are the usual scaling variables and
$L_{\mu \nu}$ is the lepton tensor. The hadronic tensor $W^{\mu \nu}$ can be
written as
\begin{equation}
2 M W^{\mu \nu}=2z \;\tr \left[2 \Phi(x)\, \g^{\mu} 
		\; 2 \Delta(z)\, \g^{\nu} \right]. \label{e:hadro}
\end{equation}

By substituting the full structure of $\Phi$ and $\Delta$ into 
Eq.~(\ref{e:hadro}) and using the resulting hadronic tensor in
Eq.~(\ref{e:cross}), 
the cross section for an unpolarized target becomes:
\begin{eqnarray} 
\lefteqn{\frac{\de \sig_{{\scriptscriptstyle U}}
	(l+\vec{H} \rightarrow l'+\vec{\vec{h}}+X)}{\de x \de z\de y}=} \nn \\
&& \qquad \frac{4 \pi \alpha^2 s}{Q^4}\;\left(1-y-\frac{y^2}{2}\right) x\; 
	 f_1(x)\left[D_1(z) + S_{h\,LL} \; D_{1LL}(z)\right],
\end{eqnarray} 
while for a transversely polarized target we obtain:
\begin{eqnarray}
\lefteqn{\frac{\de \sig_{{\scriptscriptstyle T}}
	(l+\vec{H} \rightarrow l'+\vec{\vec{h}}+X)}{\de x \de z\de y}=} \nn \\
&&  \qquad \frac{4 \pi \alpha^2 s}{Q^4}\;x\;(1-y)\;|S_{T}|\, |S_{h\,LT}|
	\sin{(\phi_{LT}+\phi_T)}
		\; h_{1}(x)  \; H_{1LT}(z), \label{e:h1}
\end{eqnarray}
where $\phi_T$ and $\phi_{LT}$ are the azimuthal angles between the scattering
 plane
and, respectively, the transverse spin of the target and the
longitudinal-transverse spin of the outgoing hadron.

Eq.~(\ref{e:h1}) shows that semi-inclusive DIS with spin-1 outgoing hadrons
allows the measurement of the transverse-spin distribution function $h_1(x)$ in
connection with the new fragmentation function $H_{1LT}(z)$. Investigation is
required to evaluate the
magnitude of this new function and of $|S_{h\,LT}|$.
The measurement requires no perpendicular
momenta to be recorded and no azimuthal asymmetries to be
computed. It is sufficient, for instance, to reverse the spin direction of a 
transversely polarized target and calculate the relative single transverse 
spin asymmetry. 

The complete list of cross sections involving transverse momentum dependent 
functions is also available \cite{io}.

%%%%%%%%%%%%%%%%%%%%%%%%%%%%%%%%%%%%%%%%%%%%%%%%%%%%%%%%%%%%%%%%%%%%%%


\begin{thebibliography}{99}

\bibitem{bls}
C.~Bourrely, E.~Leader, J.~Soffer, Phys.~Rep.~{\bf 59} (1980) 95.

\bibitem{sch}
K.~Schilling, P.~Seyboth, G.~Wolf, Nucl.~Phys.~B~{\bf 15} (1970) 397.

\bibitem{multa}
see e.g. P.J.~Mulders, R.D.~Tangerman, Nucl.~Phys.~B~{\bf 461} (1996) 197.

\bibitem{io}
A. Bacchetta, P.J. Mulders, {\em Deep inelastic leptoproduction of spin-one
hadrons}, in preparation.

\bibitem{ji}
X.~Ji, Phys.~Rev.~D~{\bf 49} (1994) 114.

\end{thebibliography}
\end{document}